\documentclass[aps,prx,twocolumn,superscriptaddress,showpacs,preprintnumbers,amsmath,amssymb]{revtex4-2}

\usepackage{graphicx}
\usepackage{amssymb}
\usepackage{amsmath}
\usepackage{mathrsfs}
\usepackage{color}
\usepackage{multirow}
\usepackage{url}
\usepackage{hyperref}

\hypersetup{
    colorlinks=true,
    linkcolor=blue,
    filecolor=blue,
    urlcolor=blue,
    citecolor=blue,
}

\begin{document}

%==============================================================================
% Authors and affiliations (REVTeX format)
%==============================================================================
\author{Zhicheng He}
\affiliation{Department of Astronomy, University of Science and Technology of China, Hefei, Anhui 230026, China}
\affiliation{School of Astronomy and Space Science, University of Science and Technology of China, Hefei, Anhui 230026, China}
\email{zcho@ustc.edu.cn}

\author{Enci Wang}
\affiliation{Department of Astronomy, University of Science and Technology of China, Hefei, Anhui 230026, China}
\affiliation{School of Astronomy and Space Science, University of Science and Technology of China, Hefei, Anhui 230026, China}

\author{Luis C. Ho}
\affiliation{Kavli Institute for Astronomy and Astrophysics, Peking University, Beijing 100871, China}

\author{Yingjie Peng}
\affiliation{Kavli Institute for Astronomy and Astrophysics, Peking University, Beijing 100871, China}

\author{Huiyuan Wang}
\affiliation{Department of Astronomy, University of Science and Technology of China, Hefei, Anhui 230026, China}
\affiliation{School of Astronomy and Space Science, University of Science and Technology of China, Hefei, Anhui 230026, China}

\author{Yong Shi}
\affiliation{Department of Astronomy, Westlake University, Hangzhou 310030, Zhejiang Province, China}

\author{Tao Wang}
\affiliation{School of Astronomy and Space Science, Nanjing University, Nanjing, Jiangsu 210093, China}

\author{Yongquan Xue}
\affiliation{Department of Astronomy, University of Science and Technology of China, Hefei, Anhui 230026, China}
\affiliation{School of Astronomy and Space Science, University of Science and Technology of China, Hefei, Anhui 230026, China}

\author{Tinggui Wang}
\affiliation{Department of Astronomy, University of Science and Technology of China, Hefei, Anhui 230026, China}
\affiliation{School of Astronomy and Space Science, University of Science and Technology of China, Hefei, Anhui 230026, China}

%==============================================================================
% Title
%==============================================================================
\title{A Step-Like Transition Underlying the Life and Death of Galaxies}

%==============================================================================
% Abstract
%==============================================================================
\begin{abstract}
Galaxy quenching is a complex and fundamental problem in astrophysics, yet a simple, unified framework connecting diverse quenching mechanisms remains lacking. Here, we derive a step-function quenching response from the fundamental principle of gas mass conservation, governed by the competition between gas-supply and gas-depletion timescales. This response also enables the gas-supply timescale to be inferred from the observed quenched fraction. Applying this method to approximately 350,000 galaxies, we find a characteristic gas-supply timescale of ~1 Gyr, reveal distinct inside-out and outside-in quenching signatures associated with supermassive black holes and dark matter halos, respectively, and identify galaxies with intermediate stellar surface mass densities as occupying a critical state for quenching and rejuvenation. Our work establishes a unified physical framework in which a simple step-function response captures how diverse physical processes regulate galaxy quenching and the resumption of star formation through their effects on the competition between gas-supply and gas-depletion timescales.
\end{abstract}



\maketitle

%==============================================================================
% Main text
%==============================================================================
\section{Introduction}
\label{sec:intro}

The transformation of star-forming galaxies into quenched systems is a central problem in galaxy evolution\cite{blanton2009,peng2010,fabian2012,somerville2015,he2026}. Despite decades of progress, galaxy quenching remains a complex phenomenon, with a variety of physical mechanisms proposed to explain the cessation of star formation. These include the heating of circumgalactic gas in massive dark matter halos\cite{birnboim2003,dekel2006}, energetic feedback from active supermassive black holes (SMBHs)\cite{ishibashi2012,zubovas2012,cicone2014,king2015,he2019,he2022,chen2022,veilleux2020,he2024,wang2024}, morphological stabilization of galactic disks through bulge growth\cite{martig2009}, and the depletion or removal of cold gas through internal and environmental processes\cite{boselli2006,rasmussen2012,peng2015,ayromlou2021,old2020,xu2026}. Although each mechanism may play an important role under different conditions, a simple and unified physical framework that connects these diverse processes to the onset of quenching has remained elusive.

Here we revisit galaxy quenching from first principles, starting with mass conservation. The ability of a galaxy to sustain star formation ultimately depends on the balance between gas depletion by star formation and the supply of its gas reservoir. This perspective provides a common physical basis for understanding how seemingly distinct quenching mechanisms may regulate the same underlying process: the availability of gas for star formation. While the gas-depletion timescale can be empirically estimated from the extended Schmidt law\cite{shi2011,shi2018}, the gas-supply timescale has remained observationally unconstrained across galaxy populations. Here we develop, to our knowledge, the only observational method that infers the distribution of gas-supply timescales from the observed quenched fraction. This framework provides a unified means of investigating when galaxies cease to sustain star formation and how different physical mechanisms drive this transition by regulating gas supply.

\section{A step-function response governed by the competition between gas-depletion and gas-supply timescales}
\label{sec:approach}

The transition from star-forming (SF) to quenched galaxies (QG) requires that the supply of cold gas can no longer compensate for its depletion (see Appendix~\ref{app:masscons} for the mass conservation equation of galaxies). We therefore formulate galaxy quenching as a competition between two characteristic timescales: the gas supply timescale, $\tau_{\rm sup}$, which replenishes the cold gas reservoir, and the gas depletion timescale, $\tau_{\rm dep}$, which removes cold gas through star formation and associated cool gas depletion processes. 
Here, we propose that the complex process of galaxy quenching can be described by a remarkably simple step function,
\begin{equation}
Q(\tau_{\rm sup},\tau_{\rm dep})=
\begin{cases}
1, & \tau_{\rm sup}>\tau_{\rm dep},\\
0, & \tau_{\rm sup}\le \tau_{\rm dep},
\end{cases}
\label{eq:quenching_function}
\end{equation}
where $Q=1$ denotes a QG and $Q=0$ denotes a SF galaxy. The observed quenched fraction therefore represents the fraction of galaxies satisfying the quenching condition. Assuming that the underlying distribution of supply timescales does not vary over the depletion timescale range considered [i.e., $p(\tau_{\rm sup}\mid\tau_{\rm dep}) = p(\tau_{\rm sup})$], the quenched fraction can be expressed as

\begin{equation}
\begin{aligned}
F_{\rm Q}(\tau_{\rm dep})
&=
\int_{0}^{\infty}
Q(\tau_{\rm sup},\tau_{\rm dep}) p(\tau_{\rm sup})\,d\tau_{\rm sup} \\
&=
\int_{\tau_{\rm dep}}^{\infty}
 p(\tau_{\rm sup})\,d\tau_{\rm sup},
\end{aligned}
\label{eq:cdf}
\end{equation}

where $p(\tau_{\rm sup})$ denotes the probability density distribution of supply timescales. Equation~(\ref{eq:cdf}) provides a direct statistical connection between the observed quenched fraction and the underlying distribution of supply times. The quenching function provides an instantaneous criterion for galaxy quenching based on the balance between gas supply and depletion. In reality, however, crossing the quenching criterion does not immediately lead to a change in the observed SF/QG classification. Instead, galaxies experience a finite response time in both the SF-to-QG and QG-to-SF directions, producing a transitional population near the quenching boundary. In a statistically steady galaxy population, galaxies undergoing the SF-to-QG and QG-to-SF transitions are expected to largely compensate each other, such that the inferred effective supply timescale distribution is only minimally affected. A detailed discussion is provided in Appendix~\ref{app:response}. The efficiency of converting gas into stars (SFE, defined as the ratio of the star formation rate to the gas mass) is observed to correlate strongly with the stellar mass surface density of galaxies, following approximately ${\rm SFE}\propto\Sigma_*^{0.5}$ \cite{shi2011,shi2018} (see the detailed formula in Appendix~\ref{app:masscons}). Since the gas depletion timescale is given by $\tau_{\rm dep}=1/{\rm SFE}$, $\Sigma_*$ serves as an observational proxy for the gas depletion timescale. This enables the competition between gas supply and gas depletion to be investigated directly using observable galaxy properties.

\begin{figure*}[t]
\centering
\includegraphics[width=0.95\textwidth]{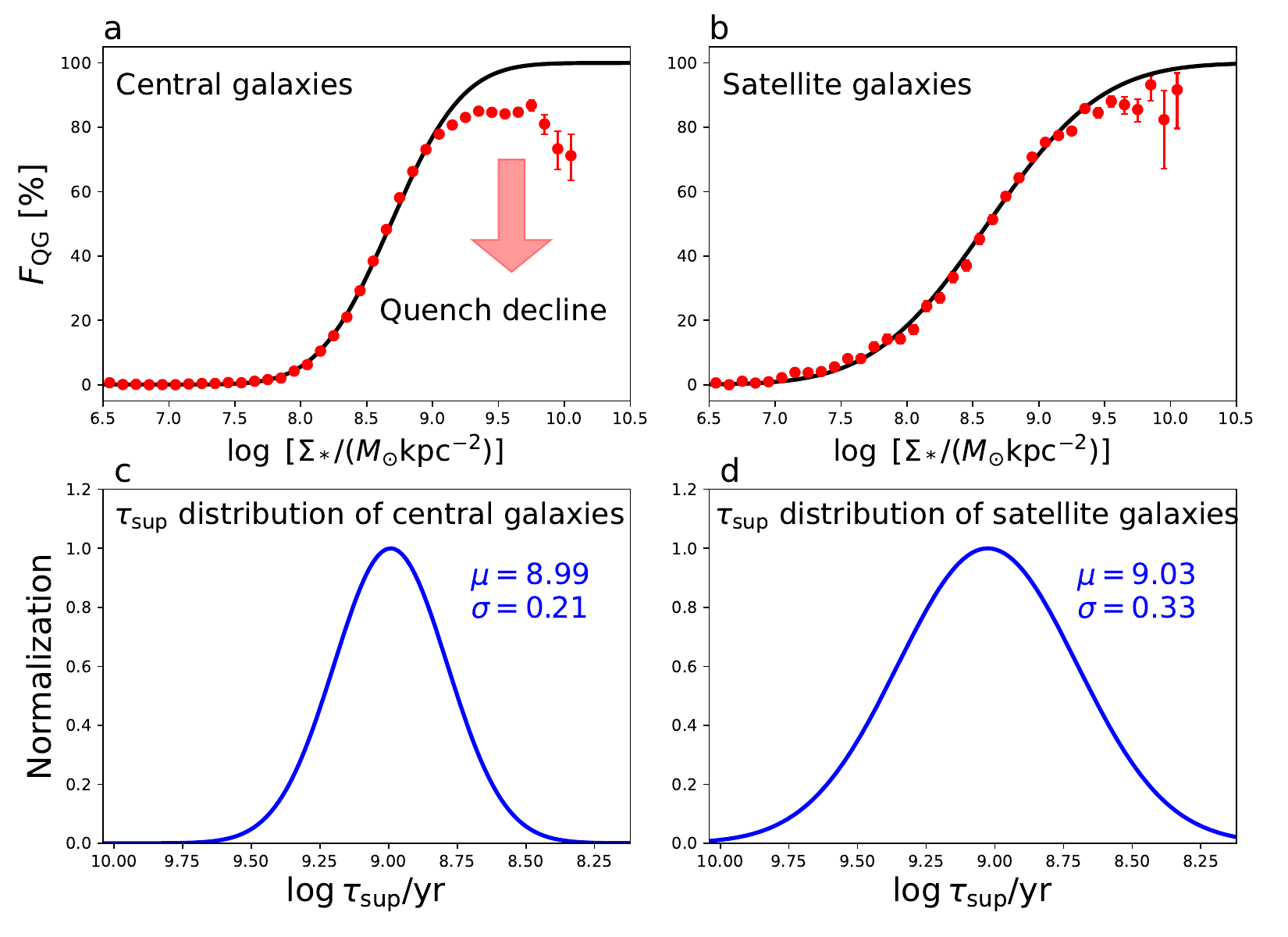}
\caption{Supply-timescale distributions inferred from the quenched fraction for a sample of 347,182 low-redshift galaxies. (a,b) Quenched fraction as a function of stellar surface density for central and satellite galaxies, respectively. At low-to-intermediate $\Sigma_*$ ($\log \Sigma_* \lesssim 9$), the solid curves show the best-fitting cumulative Gaussian models, corresponding to the cumulative distributions of supply timescales. At higher $\Sigma_*$, central galaxies exhibit a pronounced decline in the quenched fraction, whereas this trend is much weaker for satellite galaxies. (c,d) The corresponding Gaussian probability density functions inferred from the cumulative fits, representing the effective supply-timescale distributions in the low-to-intermediate-$\Sigma_*$ regime for central and satellite galaxies, respectively. The two populations have similar characteristic supply timescales, whereas the satellite population exhibits a broader distribution.}
\label{fig1}
\end{figure*}

Here we apply this method to the GSWLC--UPenn--Yang--NYU
cross-matched sample (see Appendix~\ref{app:sample} for details) containing 347,182 low-redshift galaxies. The quenching
fraction is defined as the weighted fraction of quenched galaxies in
each stellar surface density bin, where the statistical weights are
adopted to account for the survey selection function (see Appendix~\ref{app:weightedfrac} for details).
As shown in Fig.~\ref{fig1}a and b, the quenched fractions of both central and satellite galaxies increase with stellar surface density at low-to-intermediate surface densities ($\log \Sigma_* \lesssim 9$). In this regime, the observed trends are well described by cumulative Gaussian models, indicating that the
supply-timescale distribution is approximately independent of stellar surface density and can be effectively characterized by a Gaussian function. A notable feature, however, is the pronounced decline in the quenched fraction of central galaxies at higher stellar surface densities ($\log \Sigma_* \gtrsim 9$), whereas this decline is much less evident for satellite galaxies. This downturn deviates from the monotonic trend expected for a fixed supply-timescale distribution, suggesting that the supply-timescale distribution evolves with stellar surface density and, within the present framework, shifts towards shorter timescales at high $\Sigma_*$. The corresponding supply-timescale distributions inferred separately for central and satellite galaxies from the cumulative Gaussian fits are shown in Fig.~\ref{fig1}c and d. These distributions characterize the low-to-intermediate-$\Sigma_*$ regime, where the cumulative Gaussian description provides an adequate representation of the observed quenched fractions. The characteristic supply timescales of central and satellite galaxies are similar, with $\log(\mu/{\rm yr}) = 8.99$ and 9.03, respectively, whereas satellite galaxies exhibit a broader distribution, indicating a larger diversity in their supply histories. In the following analysis, we investigate whether the inferred supply-timescale distributions and the high-$\Sigma_*$ downturn are connected to black-hole growth and dark matter halos. We focus primarily on central galaxies to minimize environmental effects, while using satellite galaxies as a comparison sample to assess the role of environmental processes.

\section{Black holes and dark matter halos regulate the supply timescale in distinct ways}
\label{sec:BHhalo}

\begin{figure*}[t]
\centering
\includegraphics[width=1.0\textwidth]{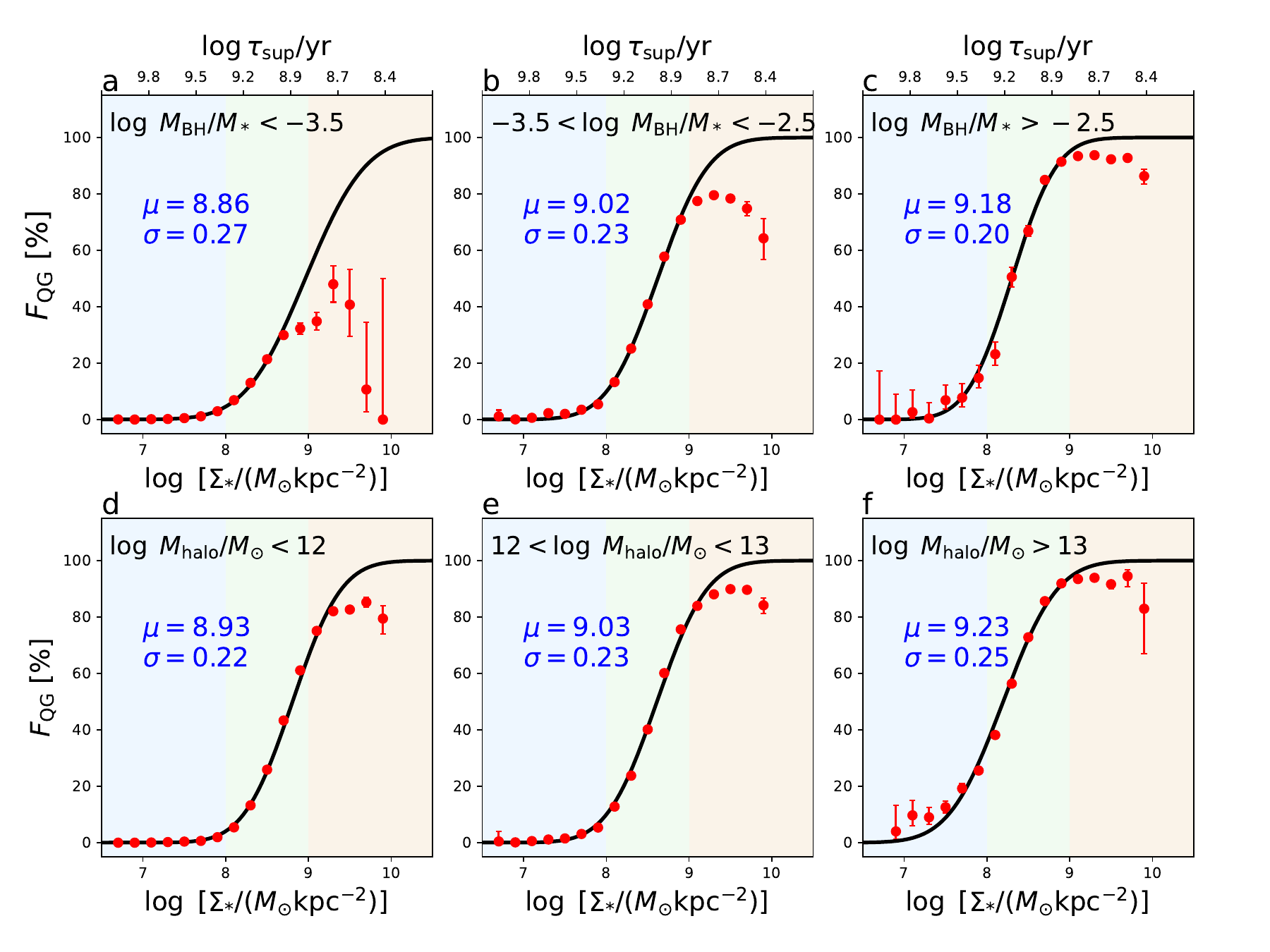}
\caption{The characteristic supply timescale increases with both $M_{\rm BH}/M_*$ and halo mass, while its distribution narrows with increasing $M_{\rm BH}/M_*$ but broadens with halo mass. (a--c) Quenched fraction as a function of stellar surface density for galaxies divided into three bins according to their black hole-to-stellar mass ratios ($M_{\rm BH}/M_*$). With increasing $M_{\rm BH}/M_*$, the characteristic supply timescale shifts toward larger values, while the width of the supply-timescale distribution decreases. The decline of the quenched fraction at high stellar surface density also becomes progressively weaker. (d--f) Same as panels (a--c), but for galaxies divided into three bins of dark matter halo mass. Increasing halo mass shifts the supply-timescale distribution toward larger values and broadens the distribution. The decrease of the quenched fraction at high stellar surface density becomes slightly weaker with increasing halo mass.}
\label{fig2}
\end{figure*}

Having established that the quenched fraction directly traces the
supply-time distribution, we now investigate how black-hole growth
and dark matter halos regulate this distribution. Figures~\ref{fig2}a--c show
the inferred supply-time distributions for galaxies divided into
three bins of the black hole-to-stellar mass ratio ($M_{\rm BH}/M_*$).
As $M_{\rm BH}/M_*$ increases, the inferred supply-time distribution
systematically shifts toward longer supply times while becoming
progressively narrower. At the same time, the decline of the quenched fraction at high $\Sigma_*$ becomes progressively weaker, indicating that black-hole growth can substantially prolong the effective gas supply timescale in compact galaxies. Figures~\ref{fig2}d--f present the same analysis for galaxies divided into three bins of dark matter halo mass. Increasing halo mass also shifts
the supply-time distribution toward longer supply times.
In contrast to the effect of black-hole growth, however, the
supply-time distribution becomes slightly broader with increasing
halo mass, while the high-density decline of the quenched fraction is
only mildly reduced.

These results demonstrate that both black-hole growth and dark matter
halos systematically regulate the supply timescale, but in distinct
ways. Black-hole growth drives galaxies toward a more uniform supply
state, whereas more massive halos introduce a broader distribution of
supply timescales while simultaneously increasing the characteristic
supply timescale. As shown in Fig.~\ref{fig:app_satellite_supply}, we repeated
the same analysis for satellite galaxies and found
qualitatively consistent trends, indicating that the effects of
black-hole growth and dark matter halo mass on the inferred
supply-timescale distribution are not limited to central galaxies.
As shown in Figs.~\ref{fig:app_morph_central} and~\ref{fig:app_morph_satellite}, we also tested the impact of
morphology-dependent black hole mass
estimators. Using separate $M_{\rm BH}$--$\sigma$ relations for
different galaxy morphologies yields qualitatively consistent results
for both central and satellite galaxies. The inferred supply-time
distributions and quenched fraction trends remain unchanged,
demonstrating that our conclusions are robust against the adopted
black hole mass calibration.

\begin{figure*}[t]
\centering
\includegraphics[width=0.8\textwidth]{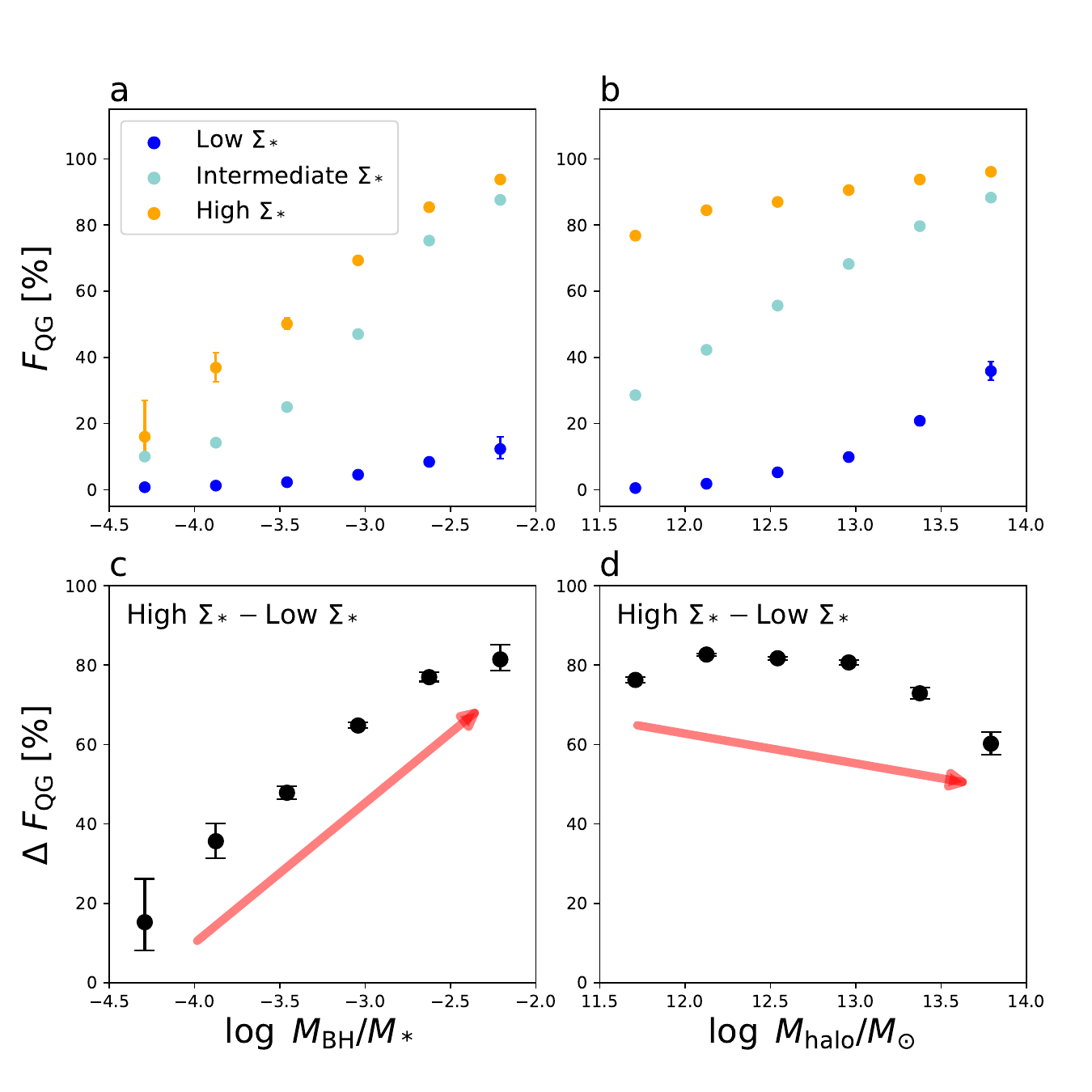}
\caption{Galaxy quenching is most sensitive to black-hole and halo regulation at intermediate stellar surface densities. (a) Quenched fraction as a function of black hole-to-stellar mass ratio ($M_{\rm BH}/M_*$) in three stellar surface density regimes: low, intermediate, and high $\Sigma_*$. The quenched fraction increases with $M_{\rm BH}/M_*$ in all surface density regimes, with different amplitudes among the three regimes. (b) Same as panel (a), but showing the dependence of the quenched fraction on dark matter halo mass. The quenched fraction also varies with halo mass, with the trends shown separately for the three $\Sigma_*$ regimes. (c) Difference between the quenched fractions at high and low stellar surface densities, $\Delta F =F_{\rm high\Sigma_*}-F_{\rm low\Sigma_*}$, as a function of $M_{\rm BH}/M_*$. The difference increases continuously with increasing black hole-to-stellar mass ratio. (d) Same as panel (c), but as a function of dark matter halo mass. The difference between the high- and low-$\Sigma_*$ quenched fractions shows a weak decreasing trend with increasing halo mass.}
\label{fig3}
\end{figure*}

\begin{figure*}[t]
\centering
\includegraphics[width=1.0\textwidth]{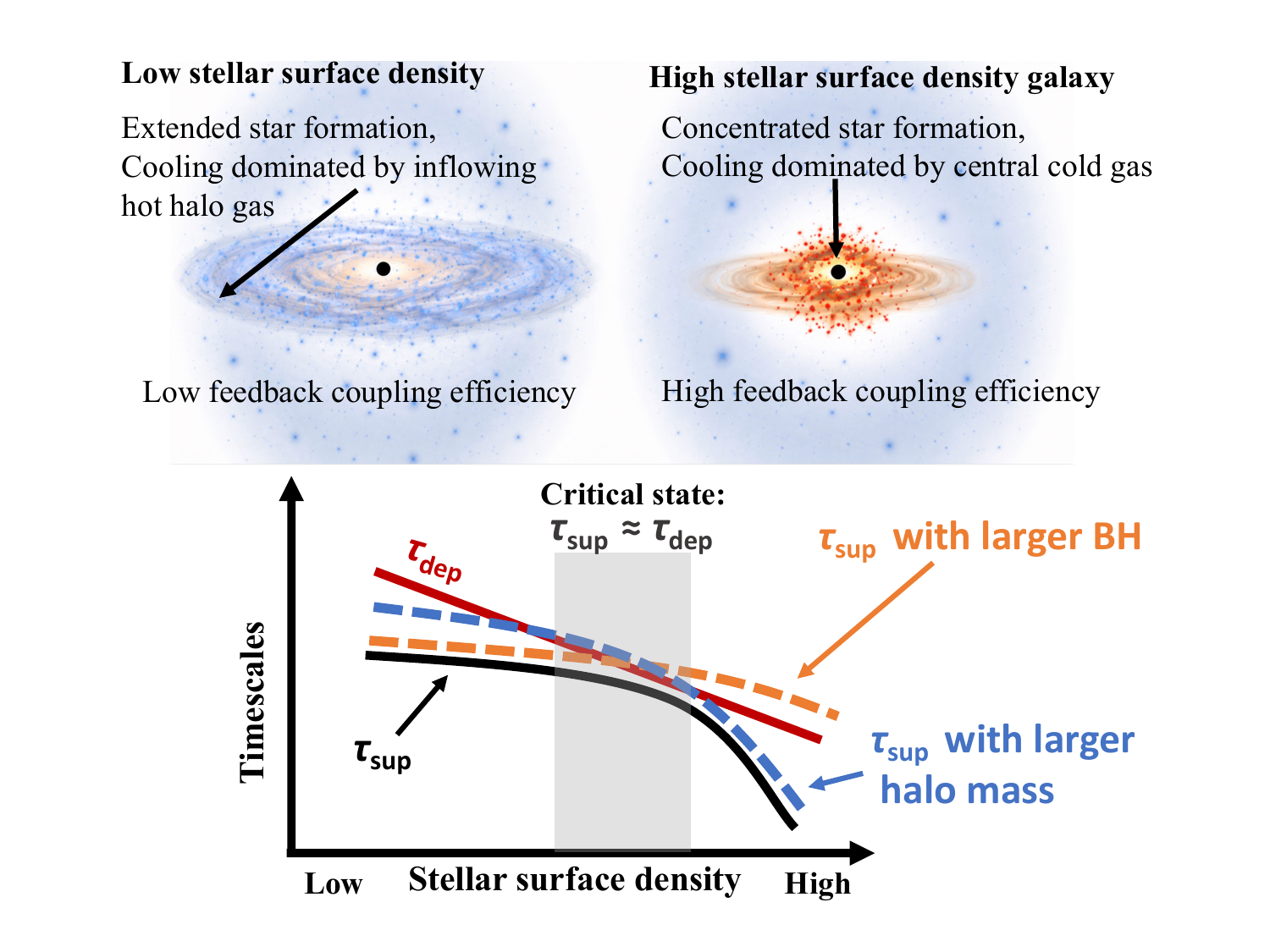}
\caption{An observationally motivated picture for galaxy quenching. At low and intermediate $\Sigma_*$, star formation is spatially extended and gas supply is primarily regulated by cooling from the hot circumgalactic halo, making the supply timescale nearly independent of $\Sigma_*$. At high $\Sigma_*$, star formation and cold gas become increasingly concentrated toward the central region, causing the supply timescale to decrease rapidly. Together with the declining depletion timescale, this naturally makes the two timescales most comparable at intermediate $\Sigma_*$ (critical state marked by gray shading), where quenching is most sensitive to changes in the supply timescale. Because SMBH feedback acts most efficiently on centrally concentrated gas, SMBH preferentially increases the supply timescale of high-$\Sigma_*$ galaxies, whereas dark matter halos exert a relatively stronger influence on low-$\Sigma_*$ galaxies.}
\label{fig4}
\end{figure*}

\section{Intermediate stellar surface density marks a critical regime for galaxy quenching}
\label{sec:critical}

Having established that both black-hole growth and dark matter halo
mass regulate the supply-timescale distribution, we now investigate
how these effects are reflected in the observed quenched fraction.
Figure~\ref{fig3} presents the dependence of the quenched fraction on the black
hole-to-stellar mass ratio ($M_{\rm BH}/M_*$) and dark matter halo
mass for galaxies with different $\Sigma_*$.

Figure~\ref{fig3}a shows that the quenched fraction increases systematically
with $M_{\rm BH}/M_*$ for galaxies spanning low, intermediate, and
high $\Sigma_*$. The dependence is strongest for
galaxies with intermediate and high $\Sigma_*$, whereas galaxies with
low $\Sigma_*$ exhibit only a weak dependence on black
hole growth. Figure~\ref{fig3}b presents the corresponding dependence on dark matter halo mass. The quenched fraction increases systematically with halo mass in all three $\Sigma_*$ bins. The increase is strongest for galaxies with intermediate $\Sigma_*$, while the low- and high-$\Sigma_*$ populations exhibit slightly weaker trends. Interestingly, the quenched fraction shows the strongest response at intermediate $\Sigma_*$ for both black-hole mass and dark matter halo mass. The physical origin of this behavior will be discussed in the following model.

To further quantify these trends, we define the difference in
quenched fraction between the high- and low-$\Sigma_*$
populations. As shown in Fig.~\ref{fig3}c, this difference increases
systematically with increasing $M_{\rm BH}/M_*$, indicating that
black-hole growth has a stronger impact on quenching in
high-$\Sigma_*$ galaxies than in low-$\Sigma_*$
galaxies. In contrast, the corresponding difference decreases
slightly with increasing dark matter halo mass (Fig.~\ref{fig3}d), suggesting
that the influence of dark matter halos is relatively stronger for
low-$\Sigma_*$ galaxies than for high-$\Sigma_*$
galaxies. Together, these results show that although both black-hole growth and dark matter halo mass have their strongest effects at intermediate $\Sigma_*$,
their relative influences reverse toward low and high $\Sigma_*$: halo regulation is relatively stronger at low $\Sigma_*$,
whereas black-hole regulation becomes relatively stronger at high $\Sigma_*$.
The physical interpretation of these distinct behaviours is discussed in
the following section. As shown in Fig.~\ref{fig:app_satellite_quenching}, we also repeated the same analysis for satellite galaxies.
While the overall trends remain qualitatively consistent with those of central galaxies, an interesting difference emerges in the dependence on dark
matter halo mass. In satellite galaxies, the quenched fraction of
high-$\Sigma_*$ galaxies exhibits a stronger dependence
on halo mass than that of low-$\Sigma_*$ galaxies, the
opposite of the trend found for central galaxies. The physical origin
of this contrasting behaviour is discussed in the following section.

\section{An observationally motivated model for galaxy quenching}
\label{sec:model}

Our results reveal a common physical picture for galaxy quenching in which the evolution of the cold-gas reservoir is governed by the competition between gas supply and depletion. The inferred supply timescale remains nearly constant across low and intermediate $\Sigma_*$, but decreases rapidly in the most compact galaxies, while the depletion timescale decreases systematically with stellar surface density, approximately as $\tau_{\rm dep}\propto\Sigma_*^{-0.5}$. These contrasting trends naturally produce a critical regime at intermediate $\Sigma_*$, where the two timescales become comparable and the quenched fraction becomes highly sensitive to changes in gas supply.

As illustrated in Fig.~\ref{fig4}, at low and intermediate $\Sigma_*$, star formation is spatially extended and gas supply is primarily regulated by the circumgalactic halo \cite{croton2006,somerville2015,tumlinson2017}, making the effective supply timescale relatively insensitive to stellar surface density. As galaxies become increasingly compact, gas and star formation become concentrated toward their central regions \cite{dekel2014,zolotov2015,tacchella2016}. The rapid decline in supply timescale at high $\Sigma_*$ therefore suggests an increasing contribution from the central cold-gas reservoir.

This picture also provides a physical interpretation of the distinct roles of dark matter halos and SMBHs. More massive halos increase the supply timescale by regulating gas accretion from the circumgalactic environment, with a relatively stronger effect in low-$\Sigma_*$ galaxies. In contrast, SMBH feedback acts preferentially in the central regions and therefore becomes increasingly important as galaxies become compact. The opposite effects of SMBH and halo growth on the width of the supply-timescale distribution further suggest that they regulate gas supply in fundamentally different ways: black-hole growth narrows the distribution, whereas halo growth slightly broadens it.

The broader supply-timescale distribution of satellite galaxies indicates more diverse gas-supply histories than those of central galaxies. Their distinct dependence on halo mass, including the stronger response of high-$\Sigma_*$ satellites, may reflect the additional role of environmental processing after infall into their current host halos, consistent with previous studies linking satellite quenching to infall time \cite{oman2013,wetzel2013,oman2016,rhee2017}. Together, these results suggest that apparently distinct quenching mechanisms act through a common physical balance between gas supply and depletion. Dark matter halos regulate gas supply on large scales, while SMBH feedback increasingly modifies the central gas supply as galaxies become compact. Galaxy quenching therefore emerges when this regulated gas supply can no longer sustain the rate of gas consumption by star formation.

Although we have focused on two of the most widely studied quenching mechanisms—halo regulation and SMBH feedback—our framework is not limited to these processes. Because it is rooted in mass conservation, it provides a general physical basis for understanding how any quenching mechanism operates through its effects on the competition between gas supply and depletion. Whether a mechanism suppresses gas accretion, removes the existing gas reservoir, or otherwise alters the availability of cold gas, its impact on star formation can be understood in terms of the resulting changes in the two timescales. This framework thus offers a common basis for investigating the diverse physical processes responsible for galaxy quenching.

Interestingly, the step-function response derived for galaxy quenching shares a conceptual similarity with the step-like response of photoionized gas to continuum variability, where the observability of ionic changes depends on the competition between the observing interval and the recombination timescale \cite{barlow1992,krolik1995,he2019}. Although the underlying physical processes differ, both systems can be described by a simple response function governed by the competition between characteristic timescales. This analogy highlights how complex physical responses can be reduced to simple threshold-like relations between timescales.

%==============================================================================
% Appendices (converted from Nature Methods)
%==============================================================================
\appendix

% Appendix figures use A1, A2, A3, ... continuous numbering
\renewcommand{\thefigure}{A\arabic{figure}}
\setcounter{figure}{0}

\section{The sample description}
\label{app:sample}

We constructed a cross-matched galaxy sample by combining the
GALEX--SDSS--WISE Legacy Catalog (GSWLC \cite{salim2007,saglia2016,salim2016}),
the UPenn PhotDec Catalog\cite{chen2012,makarov2014,meert2015,meert2016},
the Yang group catalog \cite{yang2007}, and the
New York University Value-Added Galaxy Catalog (NYU-VAGC;
\cite{blanton2005}).
These complementary catalogs provide stellar masses and star
formation rates (GSWLC), structural parameters including galaxy
morphology and effective radius (UPenn), environmental
classifications distinguishing central and satellite galaxies
\cite{yang2007}, and stellar velocity dispersions together
with the maximum observable volume ($V_{\rm max}$) (NYU-VAGC)
\cite{tinker2021, tinker2022}.
The stellar velocity dispersions are used to estimate
supermassive black hole masses through the $M_{\rm BH}$--$\sigma$
relation, while all statistical analyses are weighted by
$1/V_{\rm max}$ to correct for the flux-limited selection of the
SDSS main galaxy sample.

After applying all quality cuts, the final sample consists of
347,182 galaxies, including 251,995 central galaxies and 95,187
satellite galaxies, predominantly at redshifts $z<0.2$.

The stellar velocity dispersions provided by the NYU-VAGC are
measured within the fixed 3$^{\prime\prime}$ diameter aperture of the
SDSS spectroscopic fibers. To obtain a standardized estimate of the
central stellar velocity dispersion, we apply an aperture
correction to convert the measured values to those within one
eighth of the effective radius ($R_{\rm e}/8$), following the
prescription of \cite{cappellari2006}:
\begin{equation}
\sigma_{R_{\rm e}/8} =
\sigma_{\rm fib}
\left(\frac{R_{\rm fib}}{R_{\rm e}/8}\right)^{-0.066},
\end{equation}
where $\sigma_{\rm fib}$ is the velocity dispersion measured
within the SDSS fiber aperture, $R_{\rm fib}=1.5^{\prime\prime}$ is
fiber radius, and $R_{\rm e}$ is the galaxy effective radius.
The aperture-corrected velocity dispersions are subsequently
used to estimate supermassive black hole masses through the
$M_{\rm BH}$--$\sigma$ relation.

We derived the BH masses recommended by the review\cite{greene2020} for the full
galaxy sample,
\begin{equation}
\log \left(\frac{M_{\rm BH}}{M_\odot}\right)
= 7.87\pm0.06
+(4.55\pm 0.23)\times
\log\left(\frac{\sigma}{160~{\rm km~s^{-1}}}\right),
\end{equation}
where $\sigma$ denotes the aperture-corrected stellar velocity
dispersion within $R_{\rm e}/8$. This relation provides a
statistically estimate of black hole masses for large,
morphologically diverse galaxy samples.

\begin{figure*}[t]
\centering
\includegraphics[width=1.0\textwidth]{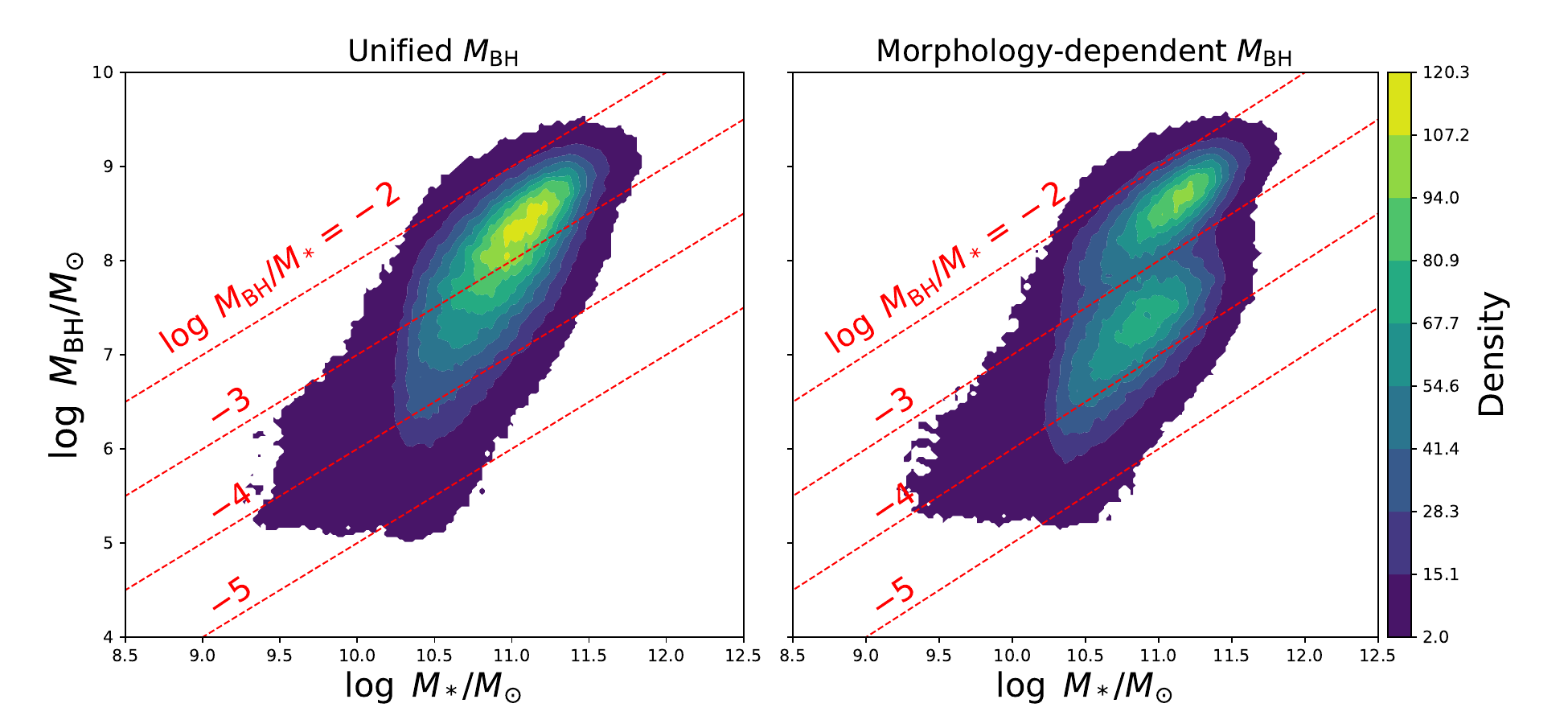}
\caption{Stellar mass versus black hole mass for the GSWLC--UPenn--Yang--NYU cross-matched galaxy sample. The left panel shows the distribution of estimated black hole masses obtained using the universal $M_{\rm BH}$--$\sigma$ relation for the full galaxy population recommended by Ref.~\cite{greene2020}. The right panel shows the corresponding relation when morphology-dependent $M_{\rm BH}$--$\sigma$ calibrations are adopted. In the latter case, galaxies are classified according to the bulge-to-total light ratio ($B/T$) used in this work, with $B/T\geq0.5$ and $B/T<0.5$ representing bulge-dominated and disk-dominated galaxies, respectively. The comparison illustrates the impact of different black hole mass calibrations on the inferred $M_{\rm BH}$--$M_*$ relation.}
\label{fig:app_bh_mass}
\end{figure*}

To assess the robustness of our results against the adopted
black hole mass calibration, we also estimate black hole masses
using morphology-dependent $M_{\rm BH}$--$\sigma$ relations.
As a simple morphological classification adopted in this work,
galaxies with $B/T\ge0.5$ are classified as bulge-dominated,
while those with $B/T<0.5$ are classified as disk-dominated.
The corresponding morphology-dependent $M_{\rm BH}$--$\sigma$
relations summarized by ref.~\cite{greene2020} are then applied to
the two subsamples. Unless otherwise stated, the scientific
conclusions presented in this work are insensitive to the adopted
black hole mass calibration.
For bulge-dominated galaxies ($B/T\ge0.5$), we adopt
\begin{equation}
\log \left(\frac{M_{\rm BH}}{M_\odot}\right)
= 8.03\pm0.06
+(4.24\pm 0.25)\times
\log\left(\frac{\sigma}{160~{\rm km~s^{-1}}}\right),
\end{equation}

whereas for disk-dominated galaxies ($B/T<0.5$), we adopt
\begin{equation}
\log \left(\frac{M_{\rm BH}}{M_\odot}\right)
= 7.44\pm0.12
+(3.61\pm 0.50)\times
\log\left(\frac{\sigma}{160~{\rm km~s^{-1}}}\right).
\end{equation}
The distribution of SMBH mass versus stellar mass for the final galaxy sample is shown in Fig.~\ref{fig:app_bh_mass}.

\section{Determination of the star-forming and quenched galaxy boundary}
\label{app:boundary}

To determine the star-forming/quenched galaxy (SF/QG) boundary, we adopted the SF/QG boundary in ref.~\cite{he2026} as the initial classification criterion. Galaxies above this boundary were used to fit the star-forming main sequence (SFMS). The fitting was performed iteratively by excluding galaxies lying more than 0.5 dex from the fitted SFMS in each iteration. The procedure was repeated until the difference in the fitted slope between two successive iterations became smaller than 0.01.

\begin{figure*}[t]
\centering
\includegraphics[width=0.8\textwidth]{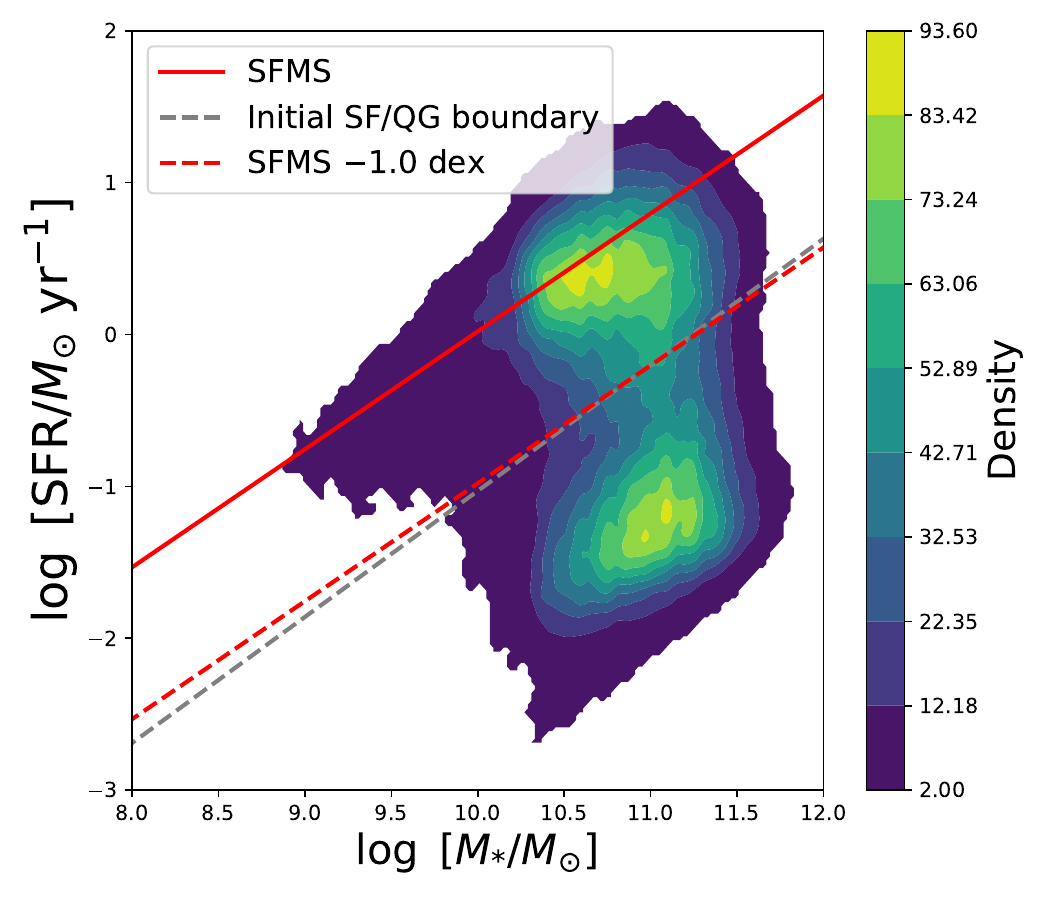}
\caption{Stellar mass--star formation rate diagram and the definition of quenched galaxies. The star formation rate (SFR) versus stellar mass for the GSWLC--UPenn--Yang--NYU cross-matched galaxy sample. The gray dashed line shows the SF/QG boundary adopted from Ref.~\cite{he2026} and used as the initial boundary for fitting the star-forming main sequence (SFMS). The solid red line represents the best-fitting SFMS obtained from our iterative fitting procedure. The red dashed line, offset by 1 dex below the best-fitting SFMS, defines the fiducial separation between star-forming and quenched galaxies.}
\label{fig:app_boundary}
\end{figure*}

As shown in Fig.~\ref{fig:app_boundary}, the resulting best-fitting SFMS is

\begin{equation}
\log \left( \frac{\mathrm{SFR}}{M_{\odot}\,\mathrm{yr}^{-1}} \right)
=
0.78\times
\log \left( \frac{M_*}{M_{\odot}} \right)
-7.74.
\end{equation}
We define the fiducial SF/QG boundary by shifting the best-fitting SFMS downward by 1.0 dex.

\section{Calculation of the weighted quenching fraction}
\label{app:weightedfrac}

The MaNGA sample is not volume complete. To account for the survey selection function, each galaxy is assigned a statistical weight, $w_i$, provided by the Pipe3D catalog. The quenching fraction is therefore calculated using weighted galaxy counts rather than the raw number of galaxies.

For a given stellar surface density bin, the weighted quenching fraction is defined as

\begin{equation}
F_{\rm Q}=
\frac{\sum\limits_{i\in {\rm QG}} w_i}
{\sum\limits_{i\in {\rm all}} w_i},
\end{equation}

where the numerator is the total weight of quenched galaxies and the denominator is the total weight of all galaxies within the same bin.

The uncertainty of the weighted quenching fraction is estimated using the Wilson score interval\cite{wilson1927}. To account for the unequal statistical weights, we first calculate the effective sample size,

\begin{equation}
N_{\rm eff}
=
\frac{\left(\sum_i w_i\right)^2}
{\sum_i w_i^2},
\end{equation}

and then evaluate the Wilson confidence interval by replacing the total sample size with $N_{\rm eff}$ and the observed fraction with the weighted quenching fraction.

\section{Mass conservation and competition between gas supply and gas depletion}
\label{app:masscons}

The evolution of the cold gas reservoir is governed by the mass conservation equation,

\begin{equation}
\frac{{\rm d}M_{\rm gas}}{{\rm d}t}
=
\dot{M}_{\rm in}
-
\dot{M}_{\rm loss}
-
{\rm SFR},
\end{equation}

where $M_{\rm gas}$ is the cold gas mass, $\dot{M}_{\rm in}$ is the gross gas inflow rate into the cold gas reservoir, including radiative cooling, recycling, and other gas supply processes, and $\dot{M}_{\rm loss}$ represents the loss rate of the cold gas reservoir arising from all physical processes, including stellar and SMBH feedback, environmental effects, and other mechanisms that reduce the net supply of cold gas. Here, $\dot{M}_{\rm loss}$ is interpreted in a broad sense, representing the net rate at which the cold gas reservoir is depleted or replenished. It includes not only the direct removal of cold gas (e.g., galactic winds, AGN-driven outflows, ram-pressure stripping, or tidal stripping), but also processes that heat the surrounding gas or suppress its cooling (e.g., stellar or AGN feedback, or virial heating), thereby reducing the effective gas cooling rate. From the perspective of the cold gas reservoir, direct gas removal is equivalent to suppressing gas cooling, since both processes decrease the supply of cold gas and therefore effectively increase the supply timescale. Likewise, processes that enhance gas cooling or promote the supply of the cold gas reservoir (i.e., positive feedback) are naturally incorporated as negative values of $\dot{M}_{\rm loss}$. In this formulation, $\dot{M}_{\rm loss}$ is not tied to any specific physical mechanism, allowing the framework to accommodate both internal and environmental processes that regulate the cold gas supply. We therefore define the net supply rate as

\begin{equation}
\dot{M}_{\rm sup}
=
\dot{M}_{\rm in}
-
\dot{M}_{\rm loss},
\end{equation}

such that the mass conservation equation becomes
\begin{equation}
\label{eq:masscons_simple}
\frac{{\rm d}M_{\rm gas}}{{\rm d}t}
=
\dot{M}_{\rm sup}
-
{\rm SFR}.
\end{equation}

The gas depletion timescale is defined as
\begin{equation}
\label{eq:taudep}
\tau_{\rm dep}
=
\frac{M_{\rm gas}}{{\rm SFR}},
\end{equation}

which characterizes the time required for star formation to consume the existing cold gas reservoir. The depletion timescale is inversely related to the star formation efficiency (SFE) and has been shown observationally to correlate tightly with the stellar surface density\cite{shi2011, shi2018}, i.e., the extended Schmidt law:
\begin{equation}
\frac{\rm SFE}{\rm yr^{-1}}=10^{-10.28}
\left(\frac{\Sigma_*}{M_{\odot} \rm pc^{-2}}\right)^{0.48},
\end{equation}
where $\Sigma_*$ is the stellar mass surface density. For a total galaxy, $\Sigma_*=M_*/(2\pi R_{\rm e}^2)$. Therefore, the depletion timescale is
\begin{equation}
\tau_{\rm dep}=\frac{1}{\rm SFE}=10^{10.28}
\left(\frac{\Sigma_*}{M_{\odot} \rm pc^{-2}}\right)^{-0.48}{\rm yr}.
\end{equation}

Consequently, in our framework, the stellar surface density serves as an observational proxy for the gas depletion timescale. Likewise, we define the supply timescale as
\begin{equation}
\label{eq:tausup}
\tau_{\rm sup}
=
\frac{M_{\rm gas}}
{\dot{M}_{\rm sup}},
\end{equation}
which represents the timescale over which the cold gas reservoir is replenished through the net supply process.

Substituting Eqs.~(\ref{eq:taudep}) and~(\ref{eq:tausup}) into Eq.~(\ref{eq:masscons_simple}) yields
\begin{equation}
\label{eq:timescale_competition}
\frac{{\rm d}M_{\rm gas}}{{\rm d}t}
=
M_{\rm gas}
\left(
\frac{1}{\tau_{\rm sup}}
-
\frac{1}{\tau_{\rm dep}}
\right).
\end{equation}

Equation~(\ref{eq:timescale_competition}) shows that the evolution of the cold gas reservoir is entirely determined by the competition between the supply timescale and the gas depletion timescale. When
$\tau_{\rm sup} < \tau_{\rm dep}$ cold gas is replenished faster than it is consumed, allowing star formation to be sustained. Conversely, when $\tau_{\rm sup} > \tau_{\rm dep}$, the cold gas reservoir is depleted faster than it is replenished, resulting in the suppression of star formation and eventually galaxy quenching. Therefore, motivated by this physical framework, we introduce the quenching function presented in the main text, which describes the probability that a galaxy is quenched as a function of the competition between the supply and depletion timescales.

\section{Effect of finite response times}
\label{app:response}

The transition between the underlying gas regulation state and the observed star formation
classification is not instantaneous. After the effective gas supply becomes insufficient, a galaxy
may still retain a fraction of its cold gas reservoir and remain on the star-forming sequence for a
finite time before being classified as quenched. Conversely, a galaxy that has recovered its gas
supply may require additional time to rebuild its cold gas reservoir and return to the
star-forming sequence.

These finite response times mainly affect galaxies close to the transition boundary between the
star-forming and quenched populations. In our analysis, the comparison between gas supply and
depletion is performed at fixed stellar surface density $\Sigma_*$, where the depletion timescale
$\tau_{\rm dep}$ varies only weakly. Therefore, the transition objects are primarily those with
effective supply timescales close to the critical value,

\begin{equation}
\tau_{\rm sup}\sim\tau_{\rm dep}.
\end{equation}

Galaxies with supply timescales significantly shorter or longer than the depletion timescale are
insensitive to such response delays, since their star formation states remain stable over the
transition timescale.

For a statistically steady galaxy population, the two types of transition objects are expected to
approximately compensate each other. The population of galaxies that have already crossed the
quenching threshold but remain observationally star-forming is balanced by the population of
galaxies that have recovered sufficient gas supply but are still observed as quenched. Thus, the
net effect on the observed quenched fraction is expected to be small:

\begin{equation}
\Delta F_{\rm QG}
=
\Delta F_{\rm SF\rightarrow QG}
-
\Delta F_{\rm QG\rightarrow SF}
\simeq 0 .
\end{equation}

Consequently, the observed quenched fraction can be used as a statistical tracer of the effective
supply timescale distribution. The inferred $\tau_{\rm sup}$ distribution should therefore be
interpreted as an effective distribution that includes the finite response between gas regulation
and the observed star formation state. Even if the two transition rates are not exactly symmetric, their contribution mainly affects galaxies near the transition boundary and introduces only a small correction to the observed quenched fraction. Therefore, the inferred effective supply timescale distribution is expected to remain robust against such finite response effects.

%==============================================================================
% Additional result figures (converted from Supplementary Figures)
%==============================================================================
\section{Additional result figures}
\label{app:additional_figs}

\begin{figure*}[t]
\centering
\includegraphics[width=1.0\textwidth]{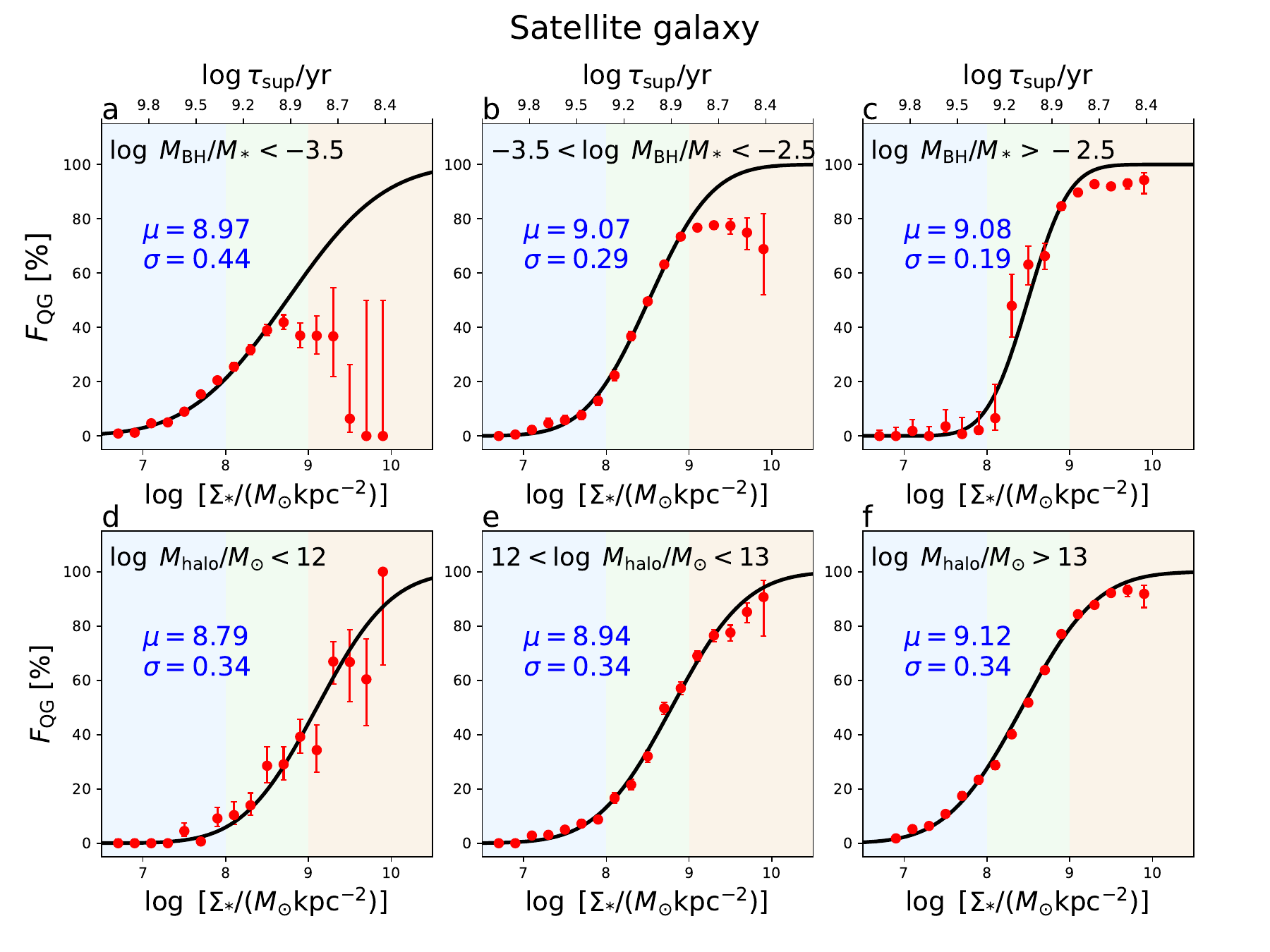}
\caption{Black hole and halo mass dependence of the supply-timescale distribution for satellite galaxies. Same as Fig.~\ref{fig2}, but for the satellite galaxy sample. (a--c) Quenched fraction as a function of stellar surface density for galaxies divided into three bins of black hole-to-stellar mass ratio ($M_{\rm BH}/M_*$). (d--f) Corresponding results for three bins of dark matter halo mass. The overall trends are qualitatively consistent with those found for central galaxies: increasing black hole mass and halo mass both shift the supply-timescale distribution toward longer supply times. Black hole growth also produces a narrower supply-timescale distribution, whereas the width of the distribution shows little dependence on halo mass for satellite galaxies.}
\label{fig:app_satellite_supply}
\end{figure*}

\begin{figure*}[t]
\centering
\includegraphics[width=1.0\textwidth]{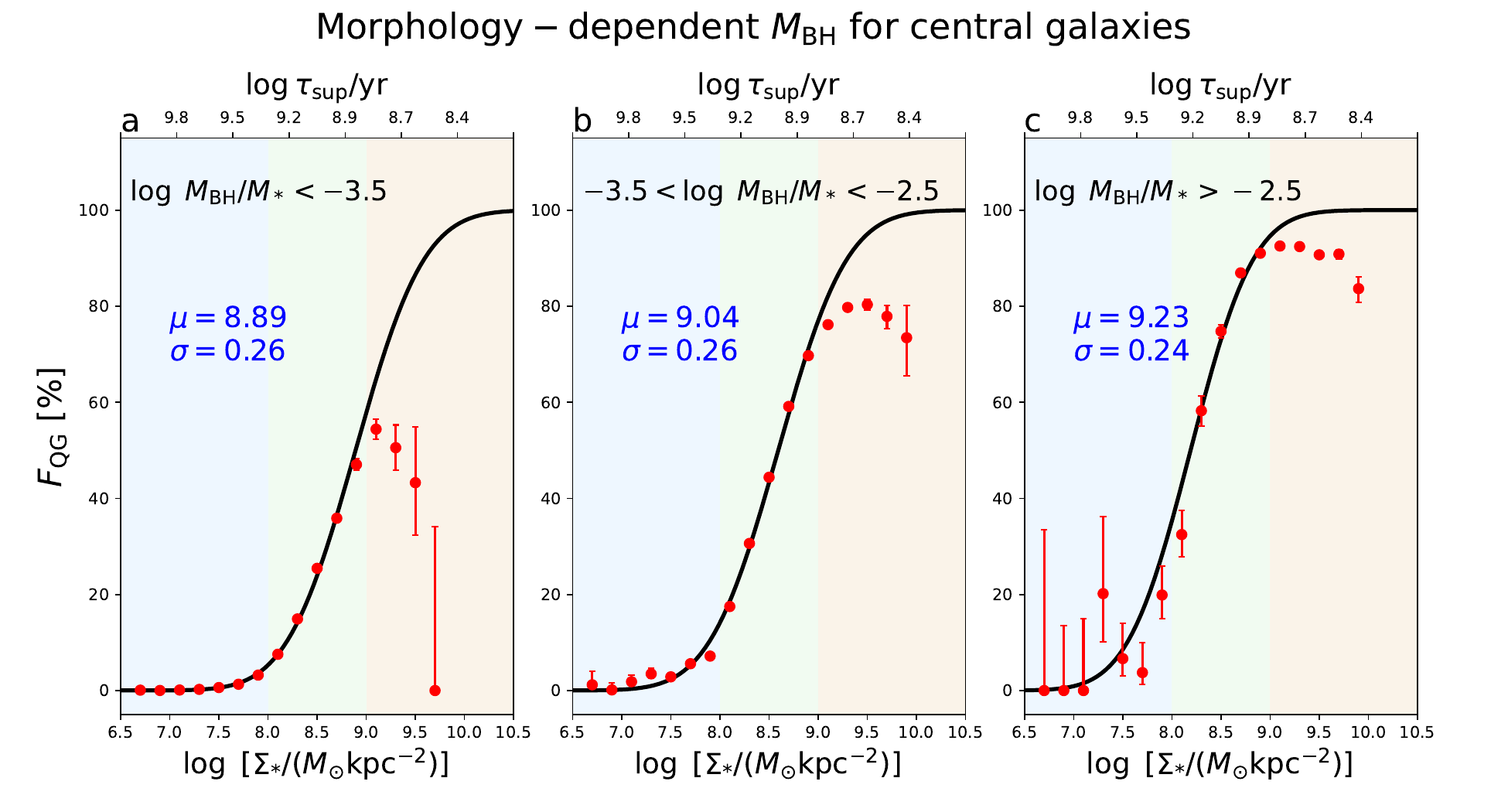}
\caption{Dependence of the supply-timescale distribution on morphology-dependent black hole mass estimates. Same as Fig.~\ref{fig2}(a--c), but using morphology-dependent black hole mass estimates instead of the unified scaling relation \cite{greene2020}. Galaxies are divided into three bins according to their black hole-to-stellar mass ratios ($M_{\rm BH}/M_*$). The overall trends remain unchanged: increasing $M_{\rm BH}/M_*$ shifts the supply-timescale distribution toward longer supply times, narrows the distribution, and weakens the decline of the quenched fraction at high stellar surface density. These results demonstrate that the inferred dependence of the supply-timescale distribution on black hole growth is robust against the adopted black hole mass estimator.}
\label{fig:app_morph_central}
\end{figure*}

\begin{figure*}[t]
\centering
\includegraphics[width=1.0\textwidth]{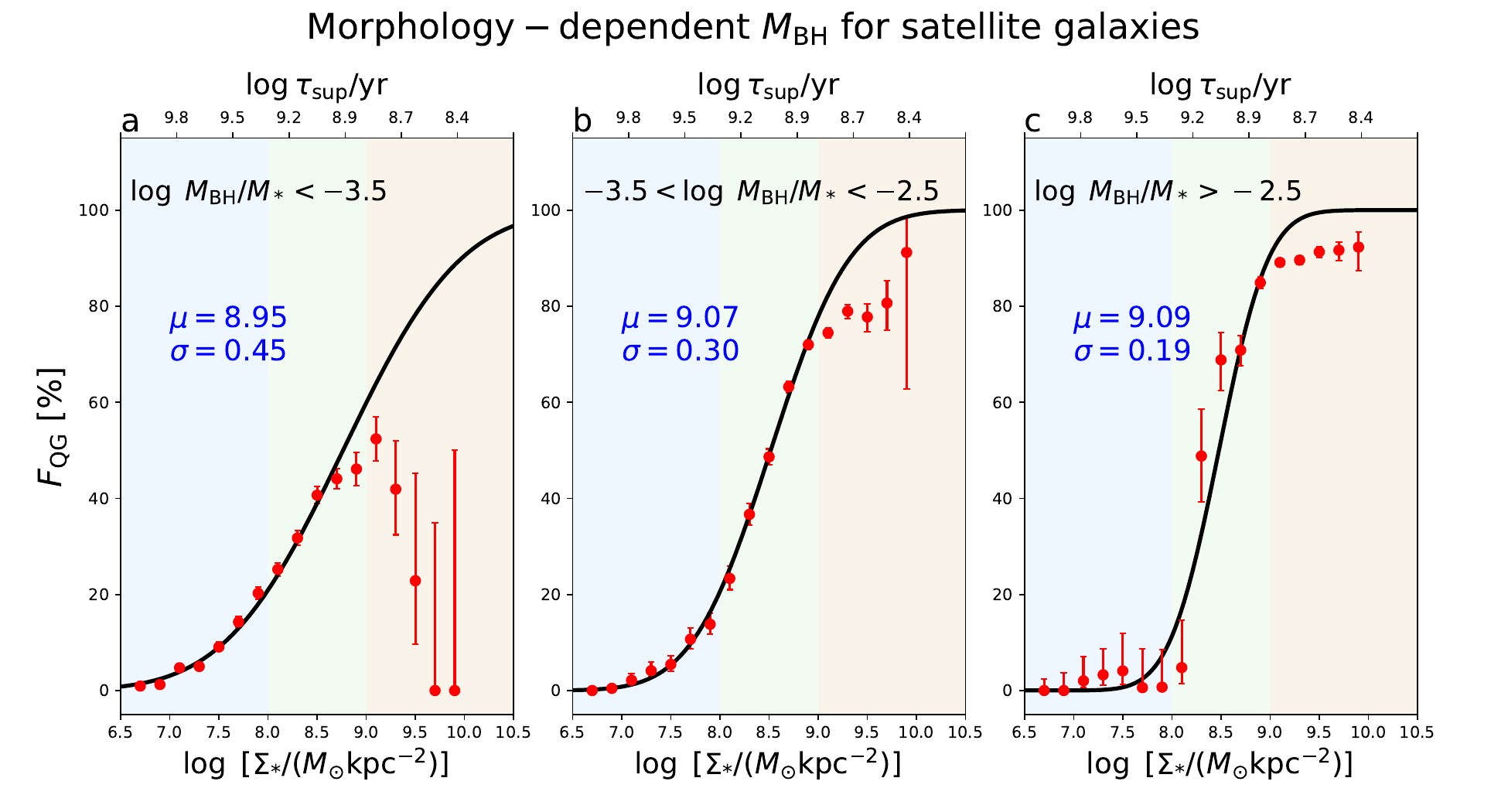}
\caption{Dependence of the supply-timescale distribution on morphology-dependent black hole mass estimates for satellite galaxies. Same as Fig.~\ref{fig:app_satellite_supply}, using morphology-dependent black hole mass estimates instead of the unified scaling relation \cite{greene2020}. Galaxies are divided into three bins according to their black hole-to-stellar mass ratios ($M_{\rm BH}/M_*$) derived from the morphology-dependent black hole mass estimators. The inferred supply-timescale distributions show the same dependence on black hole mass as obtained with the unified black hole mass estimator: larger black hole masses shift the supply-timescale distribution toward longer supply times and produce narrower distributions. These results demonstrate that the dependence of the inferred supply-timescale distribution on black hole growth is robust against the adopted black hole mass estimator for both central and satellite galaxies.}
\label{fig:app_morph_satellite}
\end{figure*}

\begin{figure*}[t]
\centering
\includegraphics[width=0.8\textwidth]{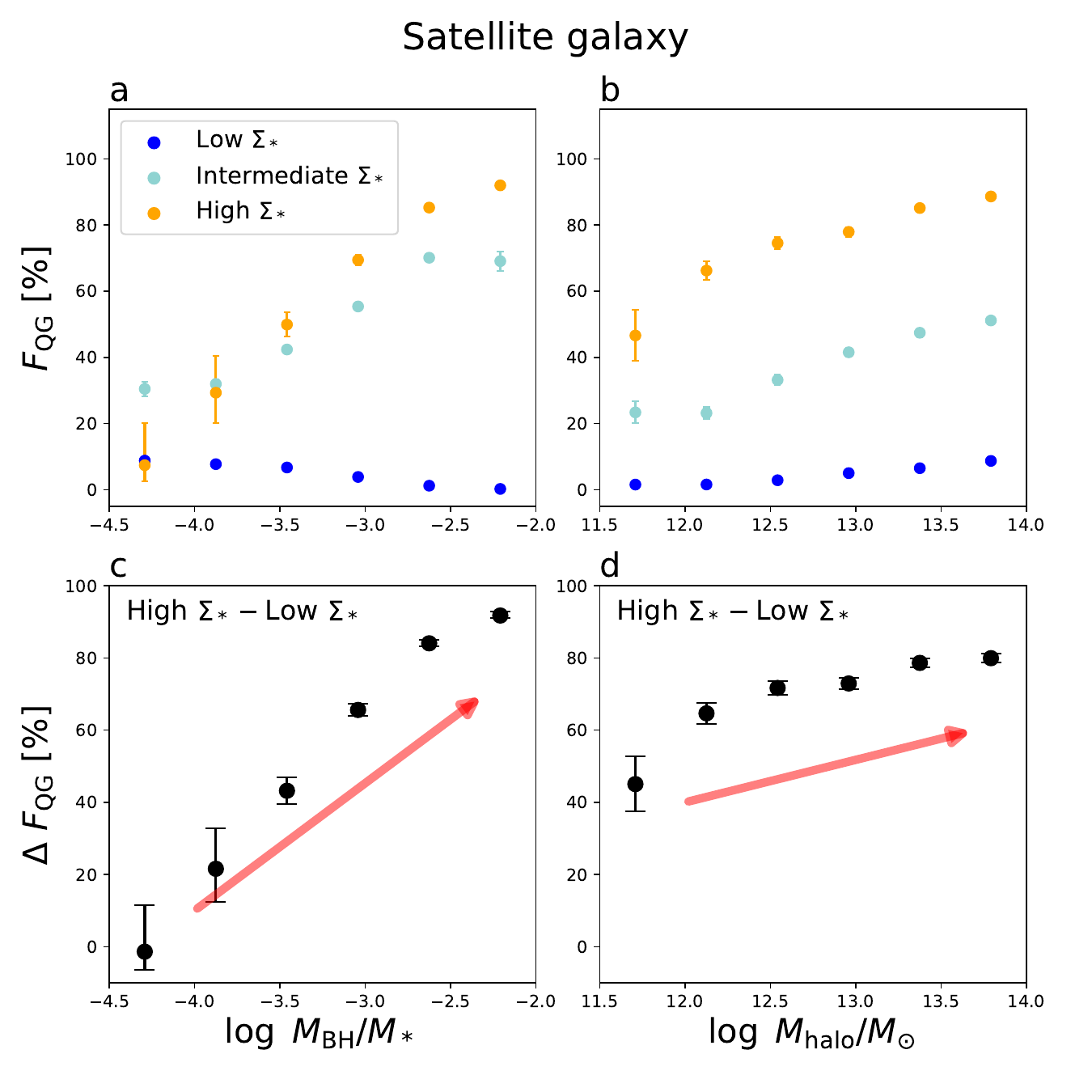}
\caption{Dependence of the quenched fraction on black hole mass ratio and halo mass for satellite galaxies. Same as Fig.~\ref{fig3}, but for the satellite galaxy sample. (a) Quenched fraction as a function of black hole-to-stellar mass ratio ($M_{\rm BH}/M_*$) for galaxies in three stellar surface density regimes. (b) Same as panel (a), but as a function of dark matter halo mass. (c) Difference between the quenched fractions at high and low $\Sigma_*$, $\Delta F = F_{\rm high\Sigma_*}-F_{\rm low\Sigma_*}$, as a function of $M_{\rm BH}/M_*$. (d) Same as panel (c), but as a function of dark matter halo mass. The overall trends are qualitatively consistent with those of central galaxies. The main difference is that the dependence on halo mass is reversed: in satellite galaxies, halo mass has a stronger influence on the quenched fraction of high-$\Sigma_*$ galaxies than on that of low-$\Sigma_*$ galaxies.}
\label{fig:app_satellite_quenching}
\end{figure*}

%==============================================================================
% Acknowledgments
%==============================================================================
\section*{Acknowledgments}
Z. C. He is supported by the National Natural Science Foundation of China (Grant Nos. 12633004, 12192220, 12192221) and the National SKA Program of China (No. 2025SKA0130100). E. Wang is supported by the National Natural Science Foundation of China, Grant Nos. 12473008. H. Y. W. is supported by the National Natural Science Foundation of China Nos. 12192224 and CAS Project for Young Scientists in Basic Research, Grant No. YSBR-062.

Funding for the Sloan Digital Sky Survey V has been provided by the Alfred P. Sloan Foundation, the Heising-Simons Foundation, the National Science Foundation, and the Participating Institutions. SDSS acknowledges support and resources from the Center for High-Performance Computing at the University of Utah. SDSS telescopes are located at Apache Point Observatory, funded by the Astrophysical Research Consortium and operated by New Mexico State University, and at Las Campanas Observatory, operated by the Carnegie Institution for Science. The SDSS web site is \url{www.sdss.org}.

SDSS is managed by the Astrophysical Research Consortium for the Participating Institutions of the SDSS Collaboration, including the Carnegie Institution for Science, Chilean National Time Allocation Committee (CNTAC) ratified researchers, Caltech, the Gotham Participation Group, Harvard University, Heidelberg University, The Flatiron Institute, The Johns Hopkins University, L'Ecole polytechnique f\'{e}d\'{e}rale de Lausanne (EPFL), Leibniz-Institut f\"{u}r Astrophysik Potsdam (AIP), Max-Planck-Institut f\"{u}r Astronomie (MPIA Heidelberg), Max-Planck-Institut f\"{u}r Extraterrestrische Physik (MPE), Nanjing University, National Astronomical Observatories of China (NAOC), New Mexico State University, The Ohio State University, Pennsylvania State University, Smithsonian Astrophysical Observatory, Space Telescope Science Institute (STScI), The Stellar Astrophysics Participation Group, Universidad Nacional Aut\'{o}noma de M\'{e}xico, University of Arizona, University of Colorado Boulder, University of Illinois at Urbana-Champaign, University of Toronto, University of Utah, University of Virginia, Yale University, and Yunnan University.

%==============================================================================
% Declarations and data availability
%==============================================================================
\section*{Data Availability}
The galaxy catalog generated in this work is publicly available through the Zenodo repository: \url{https://zenodo.org/records/21477031}. The repository contains the catalog together with documentation describing the data products and column definitions. The original survey data are available from their respective public archives.

\section*{Code Availability}
All codes used in the paper are publicly available.

\section*{Author Contributions}
Z. H. conceived the idea, performed the calculations, and wrote the manuscript. E. W., L. H., Y. P., and H. W. contributed to discussions of the idea. All coauthors discussed the results and commented on the manuscript.

\section*{Conflict of Interest}
The authors declare no competing interests.

%==============================================================================
% References
%==============================================================================
\bibliography{sn-bibliography}

\end{document}